\documentclass[aps,prb,eqsecnum,twocolumn,floatfix,supescriptaddress]{revtex4}
\usepackage{amsmath}
\usepackage{amssymb}
\usepackage[dvips]{graphicx}
\usepackage{wrapfig}
\usepackage{subfigure}
\usepackage{psfrag}

\begin{document}
\DeclareGraphicsExtensions{.eps}
\title{Quantum Hall Smectics, Sliding Symmetry and the Renormalization Group}
\author{Michael J. Lawler}
\author{Eduardo Fradkin}
\affiliation{Department of Physics, University of Illinois at Urbana-Champaign, 1110 W. Green Street, Urbana, Illinois 61801-3080, U.S.A.} 
\date{\today}

\begin{abstract}
In this paper we discuss the implication of the existence of a sliding symmetry, equivalent to the absence of a shear modulus, on the low-energy theory of the quantum hall smectic (QHS) state. We show, through renormalization group calculations, that such a symmetry causes the naive continuum approximation in the direction perpendicular to the stripes to break down through infrared divergent contributions originating from naively irrelevant operators. In particular, we show that the correct fixed point has the form of an array of sliding Luttinger liquids which is free from superficially ``irrelevant operators". Similar considerations apply to all theories with sliding symmetries.
\end{abstract}

\maketitle

\section{Introduction}
\label{sec:intro}

The two-dimensional electron gas (2DEG) in large magnetic fields exhibits a dazzling array of physical phenomena such as the quantum Hall fluids with their excitations with fractional charge and fractional statistics, which have led to a deeper understanding of Quantum Mechanics. A few years ago it was discovered that for Landau level fillings with $N\geq 2$, in addition to the expected integer quantum Hall states, the 2DEG in ultra-clean samples has highly anisotropic phases with a strongly temperature-dependent anisotropy\cite{Lilly99,Du99}.  These self-organized anisotropic states can be regarded as electronic liquid crystal phases, quantum mechanical analogs of classical liquid crystals \cite{Kivelson98,Fradkin99}.  One of these phases, the most anisotropic, is the quantum Hall smectic, or stripe phase. The quantum Hall smectic breaks translation invariance along one direction: it is metallic in one direction but insulating in the other. Another phase is the quantum nematic phase which is metallic, translationally invariant but has a finite anisotropy due to rotational symmetry breaking. A number of crystalline states may also be present. Near the middle of the $N\geq 2$ Landau level, the best presently available experimental data\cite{Cooper99,Cooper01} indicates that for these magnetic fields the 2DEG behaves as a uniform, anisotropic metallic quantum fluid, and so it is consistent with a quantum Hall nematic phase down to very low temperatures $T\gtrsim \; 20$ mK \cite{Fradkin00}.

Of the quantum liquid crystal phases, the quantum Hall smectic phase is the one that has been most extensively studied, primarily as a Hartree-Fock state \cite{Koulakov96,Fogler96,Moessner96,MacDonald00,Stanescu00,Cote00,Yi00,Barci01II,Lopatnikova01}. The quantum nematic phase is currently much less understood although significant progress has already been made on this problem \cite{Oganesyan01,Radzihovsky02}. 

Two equivalent pictures of the quantum Hall smectic state have been developed. On the one hand, semi-microscopically this state can be regarded as an array of Luttinger liquids \cite{Emery00,Vishwanath01}, similar in many ways to the stripe phases proposed in the context of high temperature superconductors and other strongly correlated systems. This picture was developed in Refs. [\onlinecite{Fradkin99,Fertig99,MacDonald00,Barci01II,Lopatnikova01}] and it is largely based on the Hartree-Fock description of the quantum Hall smectic state. In this picture, the quantum Hall smectic phase is a unidirectional charge density wave, in which the states of the partially occupied Landau level arrange themselves in a stripe-like pattern. The edges of each stripe behave as a pair of chiral Luttinger liquids with opposite chirality. This state is translationally invariant along the stripe direction and breaks translational symmetry along the perpendicular direction, as well as rotational invariance. These edge states are coupled to each other by the residual Coulomb interactions. The effective picture of the quantum Hall smectic that arises is that of an array of coupled Luttinger liquids with an infinite number of marginal operators, constrained only by the restrictions imposed by the spontaneous breaking of rotational invariance. Using Hartree-Fock methods, the effective parameters of the array of coupled chiral Luttinger liquids have been estimated, its low-lying collective modes have been studied in detail, and the stability of this phase has been investigated. This is a physically appealing and attractive picture as it captures correctly much of the basic physics of this phase. However, the Hartree-Fock state has special features, such as particle-hole symmetry for a half-filled Landau level, which are not generically present in the quantum Hall smectic even for a half-filled Landau level. (These features play an important role in the determination of the phase diagram\cite{MacDonald00,Yi00}.)

On the other hand, one can construct directly a hydrodynamic theory of the quantum Hall smectic state \cite{Barci01I,Fogler01,Cote00}. This may be done by following the consequences of the spontaneous breaking of rotational invariance and of translational invariance in one direction and by taking into consideration the dynamical dominance of the Lorentz force for electrons in a Landau level \cite{Barci01I}. The hydrodynamic theory describes essentially the same physics as a theory based on an array of coupled Luttinger liquids, and differ only by the assumption that it is possible  to regard the system as a continuous medium , {\it i.e.\/} on the validity of taking a continuum limit in the direction running perpendicular to the Luttinger liquids (or stripes). 

The quantum Hall state, as well as arrays of Luttinger liquids, have, however, an additional {\em sliding symmetry} associated with the metallic character of this state which plays a central role in the physics \cite{OHern98,Emery00,Vishwanath01,MacDonald00,Barci01I}. Here by a sliding transformation we mean a parallel rigid displacement of the charge density profile on different stripe period, which we will denote below by 
\begin{equation}
\phi(x,y) \to \phi(x,y)+f(y)
\label{eq:slide}
\end{equation}
 where $\phi(x,y)$
 denotes the phase of the charge density fluctuation along a stripe, and $y$ is the direction perpendicular to the unidirectional charge density wave. Sliding invariance then means that in this state there is no energy cost associated with this transformation. Physically this means that the charge density fluctuations do not have a shear modulus. In this sense the quantum Hall stripe is an electron smectic, an electronic liquid crystal state \cite{Kivelson98,Fradkin99}.

In this paper we discuss and compare the hydrodynamic and the coupled Luttinger liquid descriptions of the quantum Hall smectic using renormalization group (RG) methods to analyze the role of various perturbations on the effective low-energy theory. It turns out that although both descriptions are superficially equivalent ({\it i. e.\/} at the level of the symmetry), they differ substantially due to the effects of perturbations, which are non-linear and contain higher derivatives in the direction of the charge modulation. It is the main result of this paper that the continuum limit is in fact invalid and that the correct fixed ``point" which governs the quantum Hall smectic phase is given by an array of Luttinger liquids. This result is purely a consequence of keeping the sliding symmetry intact. As such, the result is more general than the context of the quantum hall smectic and extends to any continuum theory with a line of nodes in its dispersion relation.  It is therefore applicable to the hydrodynamic theory of quantum Hall nematics \cite{Radzihovsky02}, DNA Lipid Complexes \cite{OHern98} and Ring exchange Bose metals \cite{Paramekanti02}.

\section{Hydrodynamic Theory of Quantum Hall Smectics}
\label{sec:hydro}

Following Ref. [\onlinecite{Barci01I}], we begin by deriving the simplest hydrodynamic theory whose dynamics are governed by the Lorentz force and whose statics are governed solely by the broken symmetries of the state.  The Quantum Hall Smectic (QHS) is formed when the first Fourier component of the electron density in the top Landau level becomes macroscopically large along some ordering wave vector, ${\bf q}_0 \equiv q_0 {\bf {\hat q_0}}=(2\pi/\lambda)\hat{\bf y}$, which we have chosen to point along the $y$ axis. Here $\lambda$ is the period of the charge density modulation and it is identified with the stripe period.  Long wavelength deformations of the local charge density $\rho({\mathbf x})$ may then be introduced through the displacement (or phason) field $u({\bf x})$:
\begin{widetext}
\begin{equation}
  \rho({\bf x}) = \rho_0+ j_0(x)+ \textrm{Re} \bigg\{\rho_S \; 
  e^{\displaystyle{i \left({\bf q}_0\cdot {\bf x}+q_0 u(x)\right) }}+
  \rho_{2k_F} \; e^{\displaystyle{i \left(2k_F \;  {\bf {\hat z}} \times {\bf {\hat q_0}} \cdot \bf x+\phi(x)\right)}}
  \bigg\} + \ldots 
  \label{eq:density}
\end{equation}
\end{widetext}
Here $\rho_0$ is the average (uniform) density, $j_0(x)$ is the long wavelength fluctuation of the density (about the uniform value $\rho_0$) and will become the one dimensional density $\partial_x \phi$ with $\phi$ the long wavelength phase fluctuations {\em along} the stripe, $\rho_S$ is the amplitude of the unidirectional CDW (smectic) order parameter, $\varphi_S\equiv {\bf q}_0\cdot {\bf x}+q_0 u(x)$ is the phase of the complex CDW order parameter and the Goldstone boson $u$ is the displacement field which parametrizes the stripe configuration. In this paper we will only consider the physics deep in the quantum Hall smectic phase and we will thus ignore fluctuations of the amplitude of the CDW order parameter $\rho_S$, and only the phase fluctuations of the CDW order parameter, represented by the displacement field $u$, will be taken into account. Also to simplify the notation we will denote the long wavelength fluctuations of the density as $\rho$ instead of $j_0$ which we did above. Hence, we will focus on the quantum fluctuations of the Goldstone bosons of the spontaneously broken translational symmetry. In Eq. \eqref{eq:density} we have included the effects of a possible CDW ordering {\em along} the stripes. CDW ordering along the stripes is represented by a modulation of the charge density with ordering wave vector $ {\bf Q}=2k_F \hat {\bf x}$ with amplitude $\rho_{2k_F}$ and phase fluctuation $\phi$. As noted in Ref. [\onlinecite{Fradkin99}], if the phase field $\phi$ acquires rigidity {\em across} the stripes, {\it i.e.\/} along the $y$ direction, the 2DEG becomes an insulating stripe crystal. We will assume that there is no long range CDW order along the stripes in what follows.

Quantum phase fluctuations in the quantum Hall smectic phase are thus governed by an effective low-energy Quantum Hall Smectic Hamiltonian, which takes the form
\begin{align}
  {\mathcal H}_{Sm} &= \frac{1}{2}\int
  d^2x\bigg[A_1\big({\bf\nabla}\varphi_S\big)^2 
  + A_2\big({\bf\nabla}\varphi_S\big)^4\notag\\
  &\qquad +A_3\big({\bf\nabla}^2\varphi_S\big)^2\bigg],
  \label{eq:basicrot}\\
  &=\frac{1}{2}\int d^2x\bigg[C\big(\partial_y 
  u+\frac{1}{2}\big({\bf\nabla} u\big)^2\big)\notag\\
  &\qquad+\kappa_\perp\big(\partial_y  u 
  +\frac{1}{2}\big({\bf\nabla} u\big)^2\big)^2 
  + Q\big(\nabla^2
  u\big)^2\bigg],
  \label{eq:smectic} 
\end{align}   
where $\partial_y={\bf {\hat q_0}} \cdot {\bf \nabla}$ and 
$\partial_x={\bf {\hat q_0}} \times {\bf \nabla}$.
The elastic constants in Eq. \eqref{eq:basicrot} and Eq. \eqref{eq:smectic} are required to be positive (for the phase to be stable) and the constant  $C$ is chosen so that the Hamiltonian sits at a stress free minimum $\langle\partial_y u\rangle = 0$. Although this Hamiltonian, as given in Eq. \eqref{eq:smectic}, appears to break rotational invariance, a rotation of ${\bf q}_0$, which sets the direction normal to the charge modulation, does not cost any energy. Finally we note that this Hamiltonian, as  given in Eq. \eqref{eq:basicrot}, has the same form as the energy of a classical smectic fluid, where the phase $\varphi_S$ represents the height of the layered fluid\cite{Grinstein82}. 

Given the existence of the stripes, a current may also flow along them.  This current is a defining characteristic of the quantum Hall smectic: that it is a metal in one direction of space while an insulator in the other.  Thus, the fluctuations of the charge current ${\bf J}$ and long-wavelength component of the charge density $\rho$, {\it i.e.\/} the fluctuations of the uniform average charge density $\rho_0$ (which we denoted above by $j_0$), obey a continuity equation (required by charge conservation) and a geometrical constraint: 
\begin{align}
  \partial_t\rho + {\bf\nabla\cdot J} &= 0,
  \label{eq:cont}\\
  {\bf N\cdot J} &= 0,
  \label{eq:smconstraint}
\end{align}
where ${\bf N} = \frac{1}{|q_0|}{\bf \nabla}\varphi_S=\hat{q_0}+\nabla
u$ is normal to the stripes. The geometrical constraint, Eq. \eqref{eq:smconstraint}, results from freezing out the fluctuations of the amplitude of the CDW order parameter. 

We now meet the requirements of the constraint, Eq. \eqref{eq:smconstraint}, by writing the current ${\bf J}(x)$ in the form
\begin{equation}
  {\bf J}(x) \equiv {\bf J_\parallel}(x) = \sigma(x) {\bf N_\parallel}(x),
\end{equation}
where ${\bf N_\parallel} = (N_y,-N_x)$ is orthogonal to ${\bf N}$. With this definition of ${\bf J}$, Eq. \eqref{eq:cont} becomes
\begin{align}
  \partial_t\rho + D_\parallel\sigma(x) =-
  \sigma(x)\nabla\wedge\nabla u, \label{eq:1dcont}
\end{align}
where $D_\parallel \equiv {\bf N}_\parallel(x)\cdot\nabla=\varepsilon_{ij}N_j \partial_i$ and $\nabla\wedge\nabla u = \varepsilon_{ij}\partial_i\partial_j u$. 

~From this point of view, the QHS can be thought of as a collection of one-dimensional curved surfaces with singularities arising from defects in the stripe structure. These defects are dislocations whose Burgers vectors are given by the winding number of the displacement field $u$, and if there are no dislocations $\nabla \wedge \nabla u=0$. In the absence of dislocations, the charge current cannot flow at all in the direction normal to the stripes and the inter-stripe conductivity vanishes in this case. However, virtual dislocation-anti-dislocation pairs, which are high energy ``interstitial" excitations, can mediate inter-stripe tunneling processes and will govern the low frequency inter-stripe transport. A quantum Hall smectic with a non-vanishing density of interstitials, as envisioned in Ref.[\onlinecite{Radzihovsky02}], has a finite (but typically low) inter-stripe conductivity. In a stable quantum Hall smectic phase, dislocations are finite energy excitations \cite{Fradkin99}. Dislocation energies have been calculated within a Hartree-Fock state in Ref.[\onlinecite{wexler01}] who found them to be substantially large. Virtual processes in which dislocation-antidislocation pairs are created out of the ground state should give the leading contribution to finite frequency conductivity perpendicular to the stripes. Dislocation condensation is the naturally expected mechanism for a quantum phase transition to a uniform anisotropic quantum Hall nematic phase \cite{Fradkin99}.

We are now in a position to write down an imaginary time partition
function for the QHS. Including the Lorentz force and a local
density-density interaction we obtain:
\begin{equation}
\begin{split}
  {\mathcal Z} &= \int {\mathcal D}u{\mathcal D}\rho{\mathcal
  D}\sigma \delta\big(\partial_\tau\rho + D_\parallel\sigma(x) +
  \sigma(x)\nabla\wedge\nabla u\big)\times\\
  &\qquad\times\exp\big(- S[u,\rho,\sigma]\big),
  \label{eq:partrot}
\end{split}
\end{equation}
where the action $S$ is given by
\begin{equation}
  S = \int d^2xd\tau \bigg[{\bf J}\cdot{\bf A} +
  \frac{\kappa_\parallel}{2}\, \rho^2\bigg] + \int d\tau {\mathcal H}_{Sm},
\end{equation}
where ${\mathcal H}_{Sm}$ is defined in Eq.\eqref{eq:smectic}, and ${\bf A} = -B({\bf \hat{q}_0\cdot x} + u){\bf \hat{N}}_\parallel$, where $B$ is the total magnetic field,
is the Landau gauge written in the coordinates of the problem. The
choice of using the Landau gauge is natural here, although not required, given the symmetries of this state.   

In the absence of defects, $\nabla\wedge\nabla u=0$, we can simplify Eq. \eqref{eq:partrot} by making use of the approximations 
${\bf N} \approx \hat{\bf q}_0\equiv\hat{\bf y}$ and ${\bf N}_\parallel \approx \hat{\bf x}$, valid in the long wavelength limit.  
(After our calculation, we will come back and discuss this approximation.) Then, in this limit, we may now implement the
 delta-function by letting 
\begin{equation}
  \rho = \partial_x\phi \qquad \sigma = -\partial_\tau\phi,
\end{equation}
giving us the action:
\begin{multline}
  S[u,\phi] = \int d^2xd\tau 
  \bigg[-i\frac{eB}{\lambda}u\partial_\tau\phi +
  \frac{1}{2}\kappa_\parallel \big(\partial_x\phi\big)^2 +\\
  \frac{1}{2}\kappa_\perp\big(\partial_y u\big)^2 + 
  \frac{1}{2}Q\big(\partial_x^2 u\big)^2\bigg], 
  \label{eq:Suphi}
\end{multline}
where we also took the long wavelength limit of ${\mathcal H}_{Sm}$. We also notice, as was done in Ref. [\onlinecite{Barci01I}],  that the {\em phase fluctuation} of the CDW order parameter, represented by the displacement field $u$, and the charge density fluctuation represented by the field $\phi$, are conjugate variables due to the time-reversal breaking effects of the magnetic field $B$.  It is worth to note here the analogy between this canonical structure, induced by the external magnetic field, and the similar relation between the uniform local angular momentum and the local N\'eel order parameter in quantum antiferromagnets \cite{Fradkin91,Sachdev99}. 

We may further simplify the effective action of Eq. \eqref{eq:Suphi},  by letting $u = \partial_x\theta$, integrate out $\phi$ and rescale quantities according to
\begin{equation}
   \theta\rightarrow\bigg(\frac{\lambda^2}{e^2B^2}
      \frac{\kappa_\parallel}{\kappa_\perp}\bigg)^{1/4}\theta, 
   \quad y \rightarrow \sqrt{\frac{\kappa_\perp}{Q}} y,
   \quad\tau\rightarrow\frac{eB}{\lambda\sqrt{\kappa_\parallel}Q}\tau.
\end{equation}
This gives us the simplified action:
\begin{equation}
  S_0[\theta] = \frac{1}{2}\int d^2xd\tau\bigg[
  \big(\partial_\tau\theta\big)^2 +\big(\partial_y\partial_x\theta\big)^2   
  +\big(\partial_x^3\theta\big)^2\bigg],\label{eq:fp}
\end{equation}
which is the natural result of our hydrodynamic picture of this phase and was first written down independently by Barci and coworkers \cite{Barci01I} and by Fogler \cite{Fogler01}.

\section{Scaling Theories of the Quantum Hall Smectic: Perturbation Theory, RG and Dangerous Irrelevant Operators}
\label{sec:scaling}

In Ref. [\onlinecite{Barci01I}] it was shown that the quantum Hall smectic action of Eq. \eqref{eq:fp} $S_0$, as well as the action of Eq. \eqref{eq:Suphi}, are invariant  under the scale transformations: 
\begin{equation}
  \theta \rightarrow \theta,\quad x \rightarrow bx, \quad 
  y \rightarrow b^2 y, \quad  \tau \rightarrow b^3\tau.\label{eq:sclaw}
\end{equation}
and in this sense the quantum Hall smectic is at a fixed point. This scaling was used implicitly in the preceding discussion to justify our derivation. Anisotropic scaling similar to the type shown in Eq. \eqref{eq:sclaw} is common in liquid crystals and other anisotropic systems\cite{Chaikin98}. 

However, the actions given by Eq.\eqref{eq:Suphi} and Eq.\eqref{eq:fp} have the following \emph{sliding symmetry}: $\theta\rightarrow\theta + f(y)$ which follows from the charge conservation separately on each stripe (\eqref{eq:smconstraint}). Sliding symmetry has important consequences for the behavior of the system. It implies that there is no restoring force associated with sliding the charge density profiles of two stripes past each other. In other words, this state is smectic since it does not have a shear modulus. Furthermore, sliding symmetry also requires the excitation spectrum to obey
\begin{equation}
\lim_{q_x\rightarrow0} \omega({\bf q}) = 0 \label{eq:limqx}
\end{equation}
{\em for all} values of $q_y$, from small values all the way up to its natural cutoff, the inverse stripe wavelength. Thus, the low energy subspace {\em does not} automatically include {\em only} small values of $q_y$, as we have implicitly assumed above, but {\em all} values of $q_y$ must be included. Hence at small $q_x$, operators with any number of derivatives in $y$ may be important, and the validity of our continuum limit in the y-direction must be checked. On the other hand, if operators with an arbitrary number of $y$  derivatives are to be kept, the quantum Hall smectic should be better regarded as an array of coupled Luttinger liquids, essentially similar to that discussed in Ref. [\onlinecite{Emery00}] and Ref. [\onlinecite{Vishwanath01}]. Indeed this is the picture of the quantum Hall smectic advocated in Ref. [\onlinecite{Fradkin99}] and Ref. [\onlinecite{MacDonald00}]. In this picture, the quantum Hall smectic is a (infinitesimally) rotationally-invariant sliding Luttinger liquid (SLL) state which obeys the quasi-one-dimensional scaling laws
\begin{equation}
\theta \to \theta, \qquad y \to y, \qquad x \to b x, \qquad t \to bt
\label{eq:1dsclaw}
\end{equation}

In order to determine which picture is correct we will consider first the perturbative stability of the quantum Hall smectic fixed point and assume that the scaling laws of eq. \eqref{eq:sclaw} are correct. Thus we will examine the behavior of operators which are irrelevant under the scaling laws of Eq. \eqref{eq:sclaw} but marginal under the scaling laws of Eq. \eqref{eq:1dsclaw}.

The quantum Hall smectic fixed point, defined by the effective action of Eq.\eqref{eq:fp}, has the scaling laws of Eq.\eqref{eq:sclaw}, according to which there are a large number of naively irrelevant operators, typically involving powers of various derivatives of the dual field $\theta$. However, by definition, if an operator is irrelevant its contributions at low energies and long distances should be negligible. While this power counting argument is a clear and transparent criterion in conventional systems, in the presence of a sliding symmetry one has to exercise special care since in this case high momentum fluctuations may have low energy.

To explore this issue in detail we will consider now the perturbative effects of a class of naively irrelevant ({\it i. e.\/} irrelevant according to the power counting laws of Eq.\eqref{eq:sclaw}) operators.
To this effect, let us introduce a set of interactions of the form:
\begin{equation}
  S_I = \sum_{\{n,m\}=1}^\infty
     g_{n,m}\big(\partial_x^m\partial_y^n\theta\big)^4
\end{equation}
which are naively irrelevant, as implied by the tree-level $\beta$-functions
\begin{equation}
  \beta_{m,n} (\{g_{m,n}\}) \equiv -\frac{d g_{m,n}}{d\ln\Lambda} =
     \big(6-\Delta_{m,n}\big)g_{m,n}
\end{equation}
with $\Delta_{m,n} = 4(m+2n) \geq 12$ by power counting, and so are all irrelevant according to the scaling laws of Eq. \eqref{eq:sclaw}. Let us check if this power counting argument is correct and proceed to analyze the one-loop corrections introduced by these interactions.

\begin{figure}[h]
\psfrag{x}{$q_x$}
\psfrag{y}{$q_y$}
\begin{center}
\includegraphics[width=0.2\textwidth]{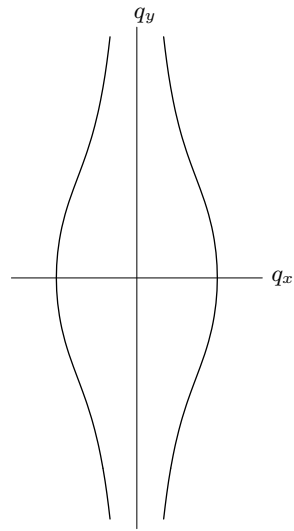}
\end{center}
\caption{The cutoff surface.}
\label{fig:cutoff}
\end{figure}

In the computation of our Feynman diagrams we will use a smooth cutoff function, consistent with the dispersion relations of the quantum Hall smectic, which has the form
\begin{equation}
e^{\displaystyle{- \left(\omega^2 + q_x^2q_y^2 + q_x^6 \right)/ \Lambda^6}}
\label{eq:cutsurf}
\end{equation}
This cutoff neither violates our small angle rotational invariance, $u\rightarrow u-\alpha x$, nor sliding symmetry.  This cutoff defines an equipotential surface depicted in Fig. \ref{fig:cutoff}. Notice the volume of phase space that the low energy modes occupy! We will have to pay particular attention to the infrared behavior of our diagrams.

The one-loop corrections to the four-point vertex function, $\Gamma^{(4)}$ typically contain the integral
\begin{equation}
  I(q) = \int  \frac{d^3k}{(2\pi)^3} 
  G^{m,n}_{\Lambda}({\bf k})G^{m,n}_{\Lambda}(q - k)
  \label{eq:I4}
\end{equation}
where $q$ is a combination of two of the four external momenta. Here we use a smooth exponential  cutoff procedure, compatible with the requirements imposed by sliding symmetry, of the form:
\begin{equation}
  G_{\Lambda}^{m,n}(k) = 
    \frac{k_x^mk_y^n}{k_0^2 + k_x^2k_y^2 + k_x^6}
        e^{-(k_0^2+k_x^2k_y^2+k_x^6)/\Lambda^6}
\end{equation}
To explore the infrared behavior of the integral of Eq. \eqref{eq:I4}, we will be interested in how well it behaves for small $q_0$ and $q_x$, though we may set $q_y$ to zero because of the the property Eq. \eqref{eq:limqx}.

To begin, let us rewrite the Green functions in the form:
\begin{equation}
  G_{\Lambda}^{m,n}(k) = k_x^mk_y^n\int_{\Lambda^{-6}}^\infty dt 
    e^{-t(k_0^2+k_x^2k_y^2+k_x^6)}
\end{equation}
After performing the integrals over $k_0$ and $k_y$ and
upon setting $q_0=0$, Eq. \eqref{eq:I4} becomes
\begin{equation}
  I \propto \int_{\Lambda^{-6}}^\infty \frac{dt_1dt_2}{\sqrt{t_1+t_2}}
   \int\frac{dk_x}{2\pi}
  \frac{k_x^m(q_x-k_x)^me^{-t_1k_x^6-t_2(q_x-k_x)^6}}{\big(t_1k_x^2 + t_2(q_x-k_x)^2\big)^{n+\frac{1}{2}}}
\end{equation}  
which is finite for finite $q_x$. 

\begin{figure}
\begin{center}
\includegraphics[width=0.6\columnwidth]{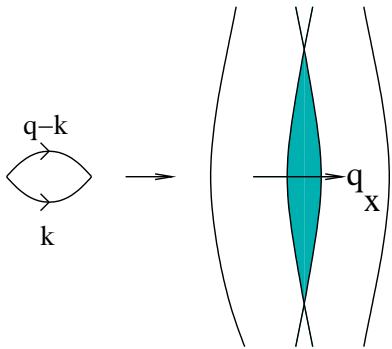}
\end{center}
\caption{Interesting ultraviolet-infrared mixing.}
\label{fig:UVIRMix}
\end{figure}

Now, let us set $q_x=0$ and perform the $k_x$ integration. The integral is:
\begin{equation}
  \int \frac{dk_x}{k_x} k_x^{2(m-n)} e^{-(t_1+t_2)k_x^6} = 
    \frac{\Gamma\big(\tfrac{1}{3}(m-n)\big)}{3\big(t_1+t_2\big)^{(m-n)/3}}
\end{equation}
where $\Gamma(z)$ is the Euler Gamma function,
and we find a strong 
{\em infrared} divergence: for $m\leq n$, $I$ diverges as we send $q_x\rightarrow0$ and the larger $n$ is than $m$, the larger the infrared divergence. Hence, to leading order, for $n > m$
\begin{equation}
  I(q_x) \propto
     \Lambda^{2m+4n-6}\bigg(\frac{\Lambda}{q_x}\bigg)^{(n-m)/3}
    \label{eq:div}
\end{equation}
while for $n=m$, we obtain a logarithmic divergence. Hence, $g_{n,m}$ is in fact {\em relevant} for $n>m$ rather than irrelevant as our naive power counting arguments would suggest. We may view this divergence as arising from an enlarging of the integration domain as $q_x\rightarrow0$. This can be seen from Fig \ref{fig:UVIRMix} as the shaded region increases as $q_x\rightarrow0$.
In other words, the low energy sector of the theory are the states with $q_x \to 0$ {\em for all} $q_y$, which follows from the sliding symmetry of the quantum Hall smectic. Therefore, operators which would be irrelevant according to the power counting rules of the scaling of Eq. \eqref{eq:sclaw} are actually {\em relevant}.

\begin{figure}[ht]
\begin{center}
\mbox{
\psfrag{z}{z}
\subfigure[r=1]{\includegraphics[width=0.23\columnwidth]{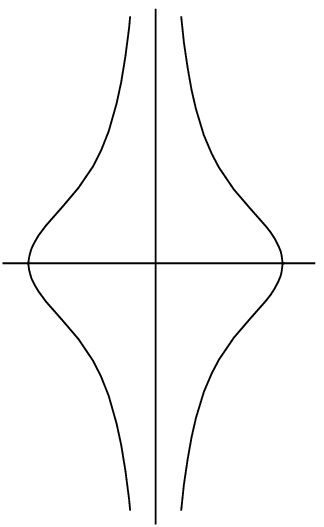}}
\subfigure[r=2]{\includegraphics[width=0.23\columnwidth]{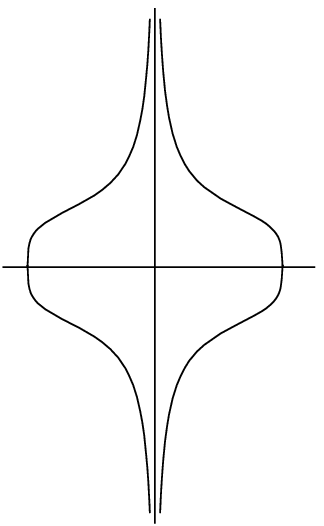}}
\subfigure[r=10]{\includegraphics[width=0.23\columnwidth]{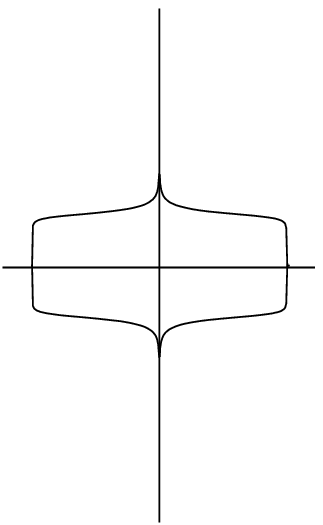}}
\subfigure[r=100]{\includegraphics[width=0.23\columnwidth]{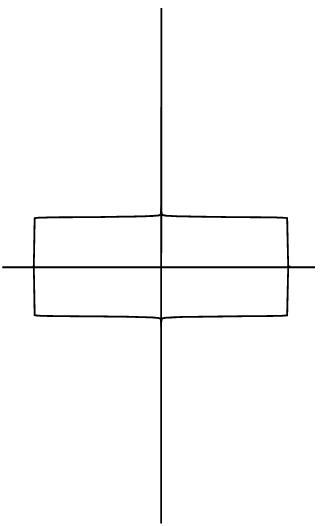}}
}
\end{center}
\caption{Cutoff shapes for $\omega_{\bf q} =
    |q_x|\sqrt{q_y^{2r}+q_x^4}=\Lambda^3$}\label{fig:cuts}
\end{figure}

The net result of this simple calculation is that we must take the consequences of Eq. \eqref{eq:limqx} seriously, and we must think of $q_y$ as not having a scaling dimension at all. To see how this arises, let us examine how to remove the infrared divergence found in Eq. \eqref{eq:div}. To that effect, let us consider the effects of adding one of the``irrelevant" operators to the action, which will now read as follows
\begin{equation}
  S_0'[\theta] = \frac{1}{2}\int d^2xdt\bigg[\big(\partial_t\theta\big)^2
    + \big(\partial_x\partial_y^r\theta\big)^2 +
    \big(\partial_x^3\theta\big)^2 \bigg],
\end{equation} 
where $r$ is a positive integer. This action is a fixed point under scaling laws similar to those of Eq. \eqref{eq:sclaw} except with $y\rightarrow b^{2/r} y$. So the larger $r$ is, the smaller is the scaling dimension of $y$ so that $y$ doesn't scale at all if we send $r\rightarrow\infty$.  Now, if we were to repeat our calculation, the cutoff surface, following the example of Eq. \eqref{eq:cutsurf}, would look like Fig \ref{fig:cuts}. Hence, the divergence, for example, that occurs for $n=m=1$ would be removed if we kept terms with y-derivatives up to $r = 2$. As expected, the addition of the relevant operator has removed the infrared divergence. In the general case, there always exists some $r$ that would remove a divergence for some $n>m$. However, for {\em any} finite value of $r$ there will be (in fact infinitely many) operators which are superficially irrelevant under the modified scaling laws but which lead to infrared divergent perturbations. Hence, in order to remove all these infrared divergences the effective action must contain {\em all} these relevant operators, {\it i.e.\/} all operators which are quadratic in the field and with two derivatives on $x$ (along the stripe direction) but with any number of derivatives in $y$, perpendicular to the stripe direction.  However for this new effective action, which now in principle contains an infinite number of coupling constants, the possible consistent scaling law are those of 
Eq. \eqref{eq:1dsclaw} according to which $y$ does not scale at all. In other words, the quantum Hall smectic has to be regarded as an array of (sliding) Luttinger liquids. This is the Sliding Luttinger liquid ``fixed point" which by construction contains an infinite number of strictly {\em marginal operators}\cite{Emery00,Vishwanath01,Fradkin99,MacDonald00}.

\section{Conclusions}
\label{sec:conclusions}

~From the discussion given above we are led to the conclusion that the continuum hydrodynamic quantum Hall smectic fixed point, characterized by the scaling laws of Eq. \eqref{eq:sclaw} is unstable, and that the correct fixed point describing the stable quantum Hall smectic phase is an array of Sliding Luttinger liquids. Like all smectics, this phase has no shear modulus \cite{DeGennes93}. However, the quantum Hall smectic and the Luttinger liquid array are electron smectics since there is no shear modulus for rigid displacements of the charge profile along the stripe direction, as required by sliding symmetry \cite{Fradkin99,Emery00}. 
Rotational invariance is spontaneously and completely broken in this stable phase of matter (see below the discussion on the stability towards crystallization). In classical three-dimensional smectics, which do not have sliding symmetry, a similar analysis shows that rotational invariance is broken mildly just by logarithmic corrections-to-scaling \cite{Grinstein82}. 

The correct scaling laws in this phase are those given in Eq. \eqref{eq:1dsclaw} where the $y$ coordinate does not scale and there is no true continuum limit in that direction. Thus, the quantum Hall smectic is best represented as an array of sliding Luttinger liquids\cite{Fradkin99} with a special form insuring invariance under infinitesimal rotations\cite{MacDonald00}.  Converting to momentum space, we may write the Hamiltonian as:
\begin{multline}
  {\mathcal H}_{d} = \frac{1}{2}\int_{-\pi/a}^{\pi/a}\frac{dq_x}{2\pi}
     \int_{-\pi/\lambda}^{\pi/\lambda}\frac{dq_y}{2\pi}\bigg[
     \kappa_\parallel(q_y)q_x^2|\hat\phi(\vec q\; )|^2 +\\
  \bigg(\kappa_\perp(q_y)\bigg(\frac{\sin(q_y\lambda/2)}{\lambda/2}\bigg)^2
      + Q(q_y)q_x^4\bigg)|\hat u(\vec q \;)|^2\bigg]
      \label{eq:Hd}
\end{multline}
where $a$ is a short distance cutoff and $\lambda$ is the stripe wavelength.
Clearly, three functions,  $\kappa_\parallel(q_y)$, $\kappa_\perp(q_y)$ and $Q(q_y)$, are now needed to be determined to fully characterize the system.  From the point of view of the QHS fixed point, these functions represent the ``dangerous irrelevant operators" discussed above.  Physically, these functions arise from forward scattering interactions coupling different stripes to each other\cite{Emery00}. The precise form of these functions has to be determined from a specific microscopic model. In Ref. [\onlinecite{MacDonald00}] and Ref. [\onlinecite{Yi00}] specific forms of these functions were calculated using Hartree-Fock approximations for the quantum Hall smectic state of the 2DEG. Except from symmetry requirements, such as invariance under infinitesimal rotations which forces the last term in Eq. \eqref{eq:Hd} to be proportional to $q_x^4$, there are few constraints on the allowed form of these strongly non-universal functions. 

It  was shown in Refs. [\onlinecite{Fradkin99,MacDonald00,Emery00,Vishwanath01}] that the functions $\kappa_\parallel(q_y)$, $\kappa_\perp(q_y)$ and $Q(q_y)$ determine the scaling dimensions of various perturbations, and hence they govern the stability of the quantum Hall smectic phase. The most important of these instabilities is the tendency to form two-dimensional crystalline states \cite{Fradkin99}. Indeed, MacDonald and Fisher \cite{MacDonald00} used a particular set of these functions, derived from a Hartree-Fock approximation of the quantum Hall smectic state assuming that particle-hole symmetry holds exactly at the center of the Landau level (as well as some monotonicity assumptions) to show that the instability towards a stripe crystal state, in which the CDW order parameters on each stripe lock to each other into a fully 2D pattern, is a marginally unstable instability. However,  Yi, Cot\'e and Fertig \cite{Yi00}, who also derived an expression for these function but without enforcing particle-hole symmetry, found instead that the quantum Hall smectic has a finite region of stability. We note that particle-hole symmetry in these systems is at best approximate and that fluctuations that are not accounted for in Hartree-Fock, such as virtual dislocation-anti-dislocation pairs, should yield finite renormalizations  of these functions without respecting particle-hole symmetry.

\begin{acknowledgments}
We are grateful to Daniel Barci, Herb Fertig, Matthew Fisher, Steve Kivelson, Tom Lubensky, Allan MacDonald and Vadim Oganesyan for many discussions on this problem. This work was supported in part by the National Science Foundation through the grant No. DMR01-32990.
\end{acknowledgments}


\begin{thebibliography}{31}
\expandafter\ifx\csname natexlab\endcsname\relax\def\natexlab#1{#1}\fi
\expandafter\ifx\csname bibnamefont\endcsname\relax
  \def\bibnamefont#1{#1}\fi
\expandafter\ifx\csname bibfnamefont\endcsname\relax
  \def\bibfnamefont#1{#1}\fi
\expandafter\ifx\csname citenamefont\endcsname\relax
  \def\citenamefont#1{#1}\fi
\expandafter\ifx\csname url\endcsname\relax
  \def\url#1{\texttt{#1}}\fi
\expandafter\ifx\csname urlprefix\endcsname\relax\def\urlprefix{URL }\fi
\providecommand{\bibinfo}[2]{#2}
\providecommand{\eprint}[2][]{\url{#2}}

\bibitem[{\citenamefont{Lilly et~al.}(1999)\citenamefont{Lilly, Cooper,
  Eisenstein, Pfeiffer, and West}}]{Lilly99}
\bibinfo{author}{\bibfnamefont{M.~P.} \bibnamefont{Lilly}},
  \bibinfo{author}{\bibfnamefont{K.~B.} \bibnamefont{Cooper}},
  \bibinfo{author}{\bibfnamefont{J.~P.} \bibnamefont{Eisenstein}},
  \bibinfo{author}{\bibfnamefont{L.~N.} \bibnamefont{Pfeiffer}},
  \bibnamefont{and} \bibinfo{author}{\bibfnamefont{K.~W.} \bibnamefont{West}},
  \bibinfo{journal}{Phys. Rev. Lett.} \textbf{\bibinfo{volume}{82}},
  \bibinfo{pages}{394} (\bibinfo{year}{1999}), \eprint{arxiv:cond-mat/9808227}.

\bibitem[{\citenamefont{Du et~al.}(1999)\citenamefont{Du, Tsui, St\"ormer,
  Pfeiffer, Baldwin, and West}}]{Du99}
\bibinfo{author}{\bibfnamefont{R.~R.} \bibnamefont{Du}},
  \bibinfo{author}{\bibfnamefont{D.~C.} \bibnamefont{Tsui}},
  \bibinfo{author}{\bibfnamefont{H.~L.} \bibnamefont{St\"ormer}},
  \bibinfo{author}{\bibfnamefont{L.~N.} \bibnamefont{Pfeiffer}},
  \bibinfo{author}{\bibfnamefont{K.~W.} \bibnamefont{Baldwin}},
  \bibnamefont{and} \bibinfo{author}{\bibfnamefont{K.~W.} \bibnamefont{West}},
  \bibinfo{journal}{Solid State Commun.} \textbf{\bibinfo{volume}{109}},
  \bibinfo{pages}{389} (\bibinfo{year}{1999}), \eprint{arXiv:cond-mat/9812025}.

\bibitem[{\citenamefont{Kivelson et~al.}(1998)\citenamefont{Kivelson, Fradkin,
  and Emery}}]{Kivelson98}
\bibinfo{author}{\bibfnamefont{S.~A.} \bibnamefont{Kivelson}},
  \bibinfo{author}{\bibfnamefont{E.}~\bibnamefont{Fradkin}}, \bibnamefont{and}
  \bibinfo{author}{\bibfnamefont{V.~J.} \bibnamefont{Emery}},
  \bibinfo{journal}{Nature} \textbf{\bibinfo{volume}{393}},
  \bibinfo{pages}{550} (\bibinfo{year}{1998}), \eprint{arXiv:cond-mat/9707327}.

\bibitem[{\citenamefont{Fradkin and Kivelson}(1999)}]{Fradkin99}
\bibinfo{author}{\bibfnamefont{E.}~\bibnamefont{Fradkin}} \bibnamefont{and}
  \bibinfo{author}{\bibfnamefont{S.~A.} \bibnamefont{Kivelson}},
  \bibinfo{journal}{Phys. Rev. B} \textbf{\bibinfo{volume}{59}},
  \bibinfo{pages}{8065} (\bibinfo{year}{1999}),
  \eprint{arXiv:cond-mat/9810151}.

\bibitem[{\citenamefont{Cooper et~al.}(1999)\citenamefont{Cooper, Lilly,
  Eisenstein, Pfeiffer, and West}}]{Cooper99}
\bibinfo{author}{\bibfnamefont{K.~B.} \bibnamefont{Cooper}},
  \bibinfo{author}{\bibfnamefont{M.~P.} \bibnamefont{Lilly}},
  \bibinfo{author}{\bibfnamefont{J.~P.} \bibnamefont{Eisenstein}},
  \bibinfo{author}{\bibfnamefont{L.~N.} \bibnamefont{Pfeiffer}},
  \bibnamefont{and} \bibinfo{author}{\bibfnamefont{K.~W.} \bibnamefont{West}},
  \bibinfo{journal}{Phys. Rev. B} \textbf{\bibinfo{volume}{60}},
  \bibinfo{pages}{R11285} (\bibinfo{year}{1999}),
  \eprint{arXiv:cond-mat/9907374}.

\bibitem[{\citenamefont{Cooper et~al.}(2001)\citenamefont{Cooper, Lilly,
  Eisenstein, Jungwirth, Pfeiffer, and West}}]{Cooper01}
\bibinfo{author}{\bibfnamefont{K.~B.} \bibnamefont{Cooper}},
  \bibinfo{author}{\bibfnamefont{M.~P.} \bibnamefont{Lilly}},
  \bibinfo{author}{\bibfnamefont{J.~P.} \bibnamefont{Eisenstein}},
  \bibinfo{author}{\bibfnamefont{T.}~\bibnamefont{Jungwirth}},
  \bibinfo{author}{\bibfnamefont{L.~N.} \bibnamefont{Pfeiffer}},
  \bibnamefont{and} \bibinfo{author}{\bibfnamefont{K.~W.} \bibnamefont{West}},
  \bibinfo{journal}{Solid State Commun.} \textbf{\bibinfo{volume}{119}},
  \bibinfo{pages}{89} (\bibinfo{year}{2001}), \eprint{arXiv:cond-mat/0104243}.

\bibitem[{\citenamefont{Fradkin et~al.}(2000)\citenamefont{Fradkin, Kivelson,
  Manousakis, and Nho}}]{Fradkin00}
\bibinfo{author}{\bibfnamefont{E.}~\bibnamefont{Fradkin}},
  \bibinfo{author}{\bibfnamefont{S.~A.} \bibnamefont{Kivelson}},
  \bibinfo{author}{\bibfnamefont{E.}~\bibnamefont{Manousakis}},
  \bibnamefont{and} \bibinfo{author}{\bibfnamefont{K.}~\bibnamefont{Nho}},
  \bibinfo{journal}{Phys. Rev. Lett.} \textbf{\bibinfo{volume}{84}},
  \bibinfo{pages}{1982} (\bibinfo{year}{2000}),
  \eprint{arXiv:cond-mat/9906064}.

\bibitem[{\citenamefont{Koulakov et~al.}(1996)\citenamefont{Koulakov, Fogler,
  and Shklovskii}}]{Koulakov96}
\bibinfo{author}{\bibfnamefont{A.~A.} \bibnamefont{Koulakov}},
  \bibinfo{author}{\bibfnamefont{M.~M.} \bibnamefont{Fogler}},
  \bibnamefont{and} \bibinfo{author}{\bibfnamefont{B.~I.}
  \bibnamefont{Shklovskii}}, \bibinfo{journal}{Phys. Rev. Lett.}
  \textbf{\bibinfo{volume}{76}}, \bibinfo{pages}{499} (\bibinfo{year}{1996}),
  \eprint{arXiv:cond-mat/9508017}.

\bibitem[{\citenamefont{Fogler et~al.}(1996)\citenamefont{Fogler, Koulakov, and
  Shklovskii}}]{Fogler96}
\bibinfo{author}{\bibfnamefont{M.~M.} \bibnamefont{Fogler}},
  \bibinfo{author}{\bibfnamefont{A.~A.} \bibnamefont{Koulakov}},
  \bibnamefont{and} \bibinfo{author}{\bibfnamefont{B.~I.}
  \bibnamefont{Shklovskii}}, \bibinfo{journal}{Phys. Rev. B}
  \textbf{\bibinfo{volume}{54}}, \bibinfo{pages}{1853} (\bibinfo{year}{1996}),
  \eprint{arXiv:cond-mat/9601110}.

\bibitem[{\citenamefont{Moessner and Chalker}(1996)}]{Moessner96}
\bibinfo{author}{\bibfnamefont{R.}~\bibnamefont{Moessner}} \bibnamefont{and}
  \bibinfo{author}{\bibfnamefont{J.~T.} \bibnamefont{Chalker}},
  \bibinfo{journal}{Phys. Rev. B} \textbf{\bibinfo{volume}{54}},
  \bibinfo{pages}{5006} (\bibinfo{year}{1996}),
  \eprint{arXiv:cond-mat/9606177}.

\bibitem[{\citenamefont{MacDonald and Fisher}(2000)}]{MacDonald00}
\bibinfo{author}{\bibfnamefont{A.~H.} \bibnamefont{MacDonald}}
  \bibnamefont{and} \bibinfo{author}{\bibfnamefont{M.~P.~A.}
  \bibnamefont{Fisher}}, \bibinfo{journal}{Phys. Rev. B}
  \textbf{\bibinfo{volume}{61}}, \bibinfo{pages}{5724} (\bibinfo{year}{2000}),
  \eprint{arXiv:cond-mat/9907278}.

\bibitem[{\citenamefont{Stanescu et~al.}(2000)\citenamefont{Stanescu, Martin,
  and Phillips}}]{Stanescu00}
\bibinfo{author}{\bibfnamefont{T.~D.} \bibnamefont{Stanescu}},
  \bibinfo{author}{\bibfnamefont{I.}~\bibnamefont{Martin}}, \bibnamefont{and}
  \bibinfo{author}{\bibfnamefont{P.}~\bibnamefont{Phillips}},
  \bibinfo{journal}{Phys. Rev. Lett.} \textbf{\bibinfo{volume}{84}},
  \bibinfo{pages}{1288} (\bibinfo{year}{2000}),
  \eprint{arXiv:cond-mat/9905116}.

\bibitem[{\citenamefont{C\^ot\'e and Fertig}(2000)}]{Cote00}
\bibinfo{author}{\bibfnamefont{R.}~\bibnamefont{C\^ot\'e}} \bibnamefont{and}
  \bibinfo{author}{\bibfnamefont{H.~A.} \bibnamefont{Fertig}},
  \bibinfo{journal}{Phys. Rev. B} \textbf{\bibinfo{volume}{62}},
  \bibinfo{pages}{1993} (\bibinfo{year}{2000}),
  \eprint{arXiv:cond-mat/0001169}.

\bibitem[{\citenamefont{Yi et~al.}(2000)\citenamefont{Yi, Fertig, and
  C\^ot\'e}}]{Yi00}
\bibinfo{author}{\bibfnamefont{H.}~\bibnamefont{Yi}},
  \bibinfo{author}{\bibfnamefont{H.~A.} \bibnamefont{Fertig}},
  \bibnamefont{and} \bibinfo{author}{\bibfnamefont{R.}~\bibnamefont{C\^ot\'e}},
  \bibinfo{journal}{Phys. Rev. Lett.} \textbf{\bibinfo{volume}{85}},
  \bibinfo{pages}{4156} (\bibinfo{year}{2000}),
  \eprint{arXiv:cond-mat/0003139}.

\bibitem[{\citenamefont{Barci and Fradkin}(2001)}]{Barci01II}
\bibinfo{author}{\bibfnamefont{D.~G.} \bibnamefont{Barci}} \bibnamefont{and}
  \bibinfo{author}{\bibfnamefont{E.}~\bibnamefont{Fradkin}},
  \bibinfo{journal}{Phys. Rev. B} \textbf{\bibinfo{volume}{65}},
  \bibinfo{pages}{245320} (\bibinfo{year}{2001}),
  \eprint{arXiv:cond-mat/0106171}.

\bibitem[{\citenamefont{Lopatnikova et~al.}(2001)\citenamefont{Lopatnikova,
  Simon, Halperin, and Wen}}]{Lopatnikova01}
\bibinfo{author}{\bibfnamefont{A.}~\bibnamefont{Lopatnikova}},
  \bibinfo{author}{\bibfnamefont{S.~H.} \bibnamefont{Simon}},
  \bibinfo{author}{\bibfnamefont{B.~I.} \bibnamefont{Halperin}},
  \bibnamefont{and} \bibinfo{author}{\bibfnamefont{X.~G.} \bibnamefont{Wen}},
  \bibinfo{journal}{Phys. Rev. B} \textbf{\bibinfo{volume}{64}},
  \bibinfo{pages}{R115301} (\bibinfo{year}{2001}),
  \eprint{arXiv:cond-mat/0105079}.

\bibitem[{\citenamefont{Oganesyan et~al.}(2001)\citenamefont{Oganesyan,
  Kivelson, and Fradkin}}]{Oganesyan01}
\bibinfo{author}{\bibfnamefont{V.}~\bibnamefont{Oganesyan}},
  \bibinfo{author}{\bibfnamefont{S.~A.} \bibnamefont{Kivelson}},
  \bibnamefont{and} \bibinfo{author}{\bibfnamefont{E.}~\bibnamefont{Fradkin}},
  \bibinfo{journal}{Phys. Rev. B} \textbf{\bibinfo{volume}{64}},
  \bibinfo{pages}{191509} (\bibinfo{year}{2001}),
  \eprint{arXiv:cond-mat/0102093}.

\bibitem[{\citenamefont{Radzihovsky and Dorsey}(2002)}]{Radzihovsky02}
\bibinfo{author}{\bibfnamefont{L.}~\bibnamefont{Radzihovsky}} \bibnamefont{and}
  \bibinfo{author}{\bibfnamefont{A.~T.} \bibnamefont{Dorsey}},
  \bibinfo{journal}{Phys. Rev. Lett.} \textbf{\bibinfo{volume}{88}},
  \bibinfo{pages}{216802} (\bibinfo{year}{2002}),
  \eprint{arXiv:cond-mat/0110083}.

\bibitem[{\citenamefont{Emery et~al.}(2000)\citenamefont{Emery, Fradkin,
  Kivelson, and Lubensky}}]{Emery00}
\bibinfo{author}{\bibfnamefont{V.~J.} \bibnamefont{Emery}},
  \bibinfo{author}{\bibfnamefont{E.}~\bibnamefont{Fradkin}},
  \bibinfo{author}{\bibfnamefont{S.~A.} \bibnamefont{Kivelson}},
  \bibnamefont{and} \bibinfo{author}{\bibfnamefont{T.~C.}
  \bibnamefont{Lubensky}}, \bibinfo{journal}{Phys. Rev. Lett.}
  \textbf{\bibinfo{volume}{85}}, \bibinfo{pages}{2160} (\bibinfo{year}{2000}),
  \eprint{arXiv:cond-mat/0001077}.

\bibitem[{\citenamefont{Vishwanath and Carpentier}(2001)}]{Vishwanath01}
\bibinfo{author}{\bibfnamefont{A.}~\bibnamefont{Vishwanath}} \bibnamefont{and}
  \bibinfo{author}{\bibfnamefont{D.}~\bibnamefont{Carpentier}},
  \bibinfo{journal}{Phys. Rev. Lett.} \textbf{\bibinfo{volume}{86}},
  \bibinfo{pages}{676} (\bibinfo{year}{2001}), \eprint{arXiv:cond-mat/0003036}.

\bibitem[{\citenamefont{Fertig}(1999)}]{Fertig99}
\bibinfo{author}{\bibfnamefont{H.~A.} \bibnamefont{Fertig}},
  \bibinfo{journal}{Phys. Rev. Lett.} \textbf{\bibinfo{volume}{82}},
  \bibinfo{pages}{3693} (\bibinfo{year}{1999}),
  \eprint{arXiv:cond-mat/9903417}.

\bibitem[{\citenamefont{Barci et~al.}(2001)\citenamefont{Barci, Fradkin,
  Kivelson, and Oganesyan}}]{Barci01I}
\bibinfo{author}{\bibfnamefont{D.~G.} \bibnamefont{Barci}},
  \bibinfo{author}{\bibfnamefont{E.}~\bibnamefont{Fradkin}},
  \bibinfo{author}{\bibfnamefont{S.~A.} \bibnamefont{Kivelson}},
  \bibnamefont{and}
  \bibinfo{author}{\bibfnamefont{V.}~\bibnamefont{Oganesyan}},
  \bibinfo{journal}{Phys. Rev. B} \textbf{\bibinfo{volume}{65}},
  \bibinfo{pages}{245319} (\bibinfo{year}{2001}),
  \eprint{arXiv:cond-mat/0105448}.

\bibitem[{\citenamefont{Fogler}(2001)}]{Fogler01}
\bibinfo{author}{\bibfnamefont{M.~M.} \bibnamefont{Fogler}}
  (\bibinfo{year}{2001}), \bibinfo{note}{unpublished; arXiv:cond-mat/0107306}.

\bibitem[{\citenamefont{O'Hern and Lubensky}(1998)}]{OHern98}
\bibinfo{author}{\bibfnamefont{C.~S.} \bibnamefont{O'Hern}} \bibnamefont{and}
  \bibinfo{author}{\bibfnamefont{T.~C.} \bibnamefont{Lubensky}},
  \bibinfo{journal}{Phys. Rev. E} \textbf{\bibinfo{volume}{58}},
  \bibinfo{pages}{5948} (\bibinfo{year}{1998}),
  \eprint{arXiv:cond-mat/9805278}.

\bibitem[{\citenamefont{Paramekanti et~al.}(2002)\citenamefont{Paramekanti,
  Balents, and Fisher}}]{Paramekanti02}
\bibinfo{author}{\bibfnamefont{A.}~\bibnamefont{Paramekanti}},
  \bibinfo{author}{\bibfnamefont{L.}~\bibnamefont{Balents}}, \bibnamefont{and}
  \bibinfo{author}{\bibfnamefont{M.~P.~A.} \bibnamefont{Fisher}},
  \bibinfo{journal}{Phys. Rev. B} \textbf{\bibinfo{volume}{66}},
  \bibinfo{pages}{054526} (\bibinfo{year}{2002}),
  \eprint{arXiv:cond-mat/0203171}.

\bibitem[{\citenamefont{Grinstein and Pelcovits}(1982)}]{Grinstein82}
\bibinfo{author}{\bibfnamefont{G.}~\bibnamefont{Grinstein}} \bibnamefont{and}
  \bibinfo{author}{\bibfnamefont{R.~A.} \bibnamefont{Pelcovits}},
  \bibinfo{journal}{Phys. Rev. A} \textbf{\bibinfo{volume}{26}},
  \bibinfo{pages}{915} (\bibinfo{year}{1982}).

\bibitem[{\citenamefont{Wexler and Dorsey}(2001)}]{wexler01}
\bibinfo{author}{\bibfnamefont{C.}~\bibnamefont{Wexler}} \bibnamefont{and}
  \bibinfo{author}{\bibfnamefont{A.~T.} \bibnamefont{Dorsey}},
  \bibinfo{journal}{Phys. Rev. B} \textbf{\bibinfo{volume}{64}},
  \bibinfo{pages}{R115312} (\bibinfo{year}{2001}),
  \eprint{arXiv:cond-mat/0009096}.

\bibitem[{\citenamefont{Fradkin}(1991)}]{Fradkin91}
\bibinfo{author}{\bibfnamefont{E.}~\bibnamefont{Fradkin}},
  \emph{\bibinfo{title}{Field Theories of Condensed Matter Systems}}
  (\bibinfo{publisher}{Addison-Wesley}, \bibinfo{address}{Massachusetts},
  \bibinfo{year}{1991}).

\bibitem[{\citenamefont{Sachdev}(1999)}]{Sachdev99}
\bibinfo{author}{\bibfnamefont{S.}~\bibnamefont{Sachdev}},
  \emph{\bibinfo{title}{Quantum Phase Transitions}}
  (\bibinfo{publisher}{Cambridge University Press},
  \bibinfo{address}{Cambridge, UK}, \bibinfo{year}{1999}).

\bibitem[{\citenamefont{Chaikin and Lubensky}(1998)}]{Chaikin98}
\bibinfo{author}{\bibfnamefont{P.~M.} \bibnamefont{Chaikin}} \bibnamefont{and}
  \bibinfo{author}{\bibfnamefont{T.~C.} \bibnamefont{Lubensky}},
  \emph{\bibinfo{title}{Principles of Condensed Matter Physics}}
  (\bibinfo{publisher}{Cambridge University Press},
  \bibinfo{address}{Cambridge, UK}, \bibinfo{year}{1998}).

\bibitem[{\citenamefont{Gennes and Prost}(1993)}]{DeGennes93}
\bibinfo{author}{\bibfnamefont{P.~G.~D.} \bibnamefont{Gennes}}
  \bibnamefont{and} \bibinfo{author}{\bibfnamefont{J.}~\bibnamefont{Prost}},
  \emph{\bibinfo{title}{The Physics of Liquid Crystals, 2$^{\text{nd}}$ Ed.}}
  (\bibinfo{publisher}{Oxford Science Publications}, \bibinfo{address}{Oxford,
  UK}, \bibinfo{year}{1993}).

\end{thebibliography}

\end{document}